\documentclass[10pt,a4paper]{article}
\usepackage{graphicx}
\usepackage{subfigure}
\usepackage{amsmath,amsthm}
\usepackage{amssymb,amsfonts}
\usepackage{authblk}
\usepackage{url}
\usepackage{hyperref}

\newtheorem{definition}{\textbf{Definition}}

\title{Two-Sample Test with Copula Entropy}
\author{Jian MA\thanks{Email: majian@hitachi.cn}}
\affil{Hitachi China Research Laboratory}
\date{}

\begin{document}

\maketitle

\begin{abstract}
	\noindent
	In this paper we propose a two-sample test based on copula entropy (CE). The proposed test statistic is defined as the difference between the CEs of the null hypothesis and the alternative. The estimator of the test statistic is proposed with the non-parametric estimator of CE, which is non-parametric and hyperparameter-free. Simulation experiments verified the effectiveness of the proposed test and compared it with three other multivariate non-parametric two-sample tests on the simulated bi-variate normal or bi-variate Gaussian copula data. Experimental results show that the proposed test works well on all three simulations and presents competitive or better performance than the other three tests.
\end{abstract}

\section{Introduction}
Two-sample test is one of the most common problems in statistics. It tests with a statistic the hypothesis that two samples are from the same probability distribution or not. There are many existing methods for this, such as the two-sample T-test, Kolmogorov-Smirnov test, Pearson's chi-squared test, Kernel-based test, etc. However, each test has its own drawbacks. For example, two-sample T-test requires the normal assumption. If the samples are non-normal, one needs to perform a non-parametric test. The Kolmogorov-Smirnov test \cite{Kolmogorov1933} is such a non-parametric test. However, it works only for uni-variate cases. The other well-known tests are kernel two-sample test \cite{Gretton2012,Shekhar2022} which defines its test statistic with kernel tricks, and the test based on energy statistics \cite{Szekely2004}. The drawback of the kernel-based test is its need for hyperparameter tuning. Both kernel-based and energy statistics based tests have high computational complexity. Recently, a test based on mutual information was proposed, which shows good performance over the other methods \cite{Guha2014}. Copula is the probabilistic theory for representing dependence and a two-sample test for tail copulas was proposed recently for financial applications \cite{Can2023}.

Copula Entropy (CE) is a kind of Shannon entropy defined with copula function \cite{Ma2011}. It has been proven that mutual information is equivalent to negative CE. CE is a multivariate measure of statistical independence with several good properties, such as being symmetric, non-positive (0 iff independent), invariant to monotonic transformation, and particularly, equivalent to correlation coefficient in Gaussian cases. Previously, CE has been applied to multivariate normality test, in which a test statistic based on CE is proposed \cite{Ma2022}.

In this paper, we propose a two-sample test with CE. The test statistic is defined as the difference between the CEs of the null hypothesis and the alternative. This is different from the previously proposed test based on MI, which only considers the null hypothesis in its test statistic \cite{Guha2014}. Another merit of our test is that the proposed test statistic can be easily estimated from data with the non-parametric estimator of CE, which makes the proposed test both non-parametric and hyperparameter-free. We evaluated the proposed test and compared it with three other multivariate non-parametric two-sample tests, including the MI-based test, the kernel-based test, and the energy statistics-based test, on simulated bi-variate normal or bi-variate Gaussian copula data.

This paper is organized as follows: Section \ref{sec:ce} introduces the basic theory of CE; the proposed test statistic and its estimator are presented in Section \ref{sec:test}; simulation experiments will be presented in Section \ref{sec:sim}; and experimental results will be presented in Section \ref{sec:results}; some discussion are presented in Section \ref{sec:discussion}; Section \ref{sec:con} concludes the paper.

\section{Copula Entropy}
\label{sec:ce}
Copula theory is a probabilistic theory on representation of multivariate dependence \cite{nelsen2007,joe2014}. According to Sklar's theorem \cite{sklar1959}, any multivariate density function can be represented as a product of its marginals and copula density function (cdf) which represents dependence structure among random variables. 

With copula theory, Ma and Sun \cite{Ma2011} defined a new mathematical concept, named Copula Entropy, as follows:
\begin{definition}[Copula Entropy]
	Let $\mathbf{X}$ be random variables with marginals $\mathbf{u}$ and copula density function $c$. The CE of $\mathbf{X}$ is defined as
	\begin{equation}
	H_c(\mathbf{x})=-\int_{\mathbf{u}}{c(\mathbf{u})\log c(\mathbf{u})d\mathbf{u}}.
	\label{eq:ce}
	\end{equation}	
\end{definition}

A non-parametric estimator of CE was also proposed in \cite{Ma2011}, which composed of two simple steps:
\begin{enumerate}
	\item estimating empirical copula density function;
	\item estimating the entropy of the estimated empirical copula density.
\end{enumerate}
The empirical copula density in the first step can be easily derived with rank statistic. With the estimated empirical copula density, the second step is essentially a problem of entropy estimation, which can be tackled with the KSG estimation method \cite{Kraskov2004}. In this way, a non-parametric method for estimating CE was proposed \cite{Ma2011}.

\section{Two-sample test with CE}
\label{sec:test}
Given two samples $\mathbf{X}_1 =\{X_{11},\cdots,X_{1m}\} \sim  P_1, \mathbf{X}_2 =\{X_{21},\cdots,X_{2n}\}\sim P_2$, the null hypothesis for two sample test is
\begin{equation}
	H_0: P_1 = P_2,
\end{equation}
and the alternative is
\begin{equation}
H_1: P_1 \neq P_2.
\end{equation}
where $\mathbf{X}_1,\mathbf{X}_2 \in R^d$ and $P_1,P_2$ are the corresponding probability distribution functions.

Let $\mathbf{X} = (\mathbf{X}_1, \mathbf{X}_2)$ and $Y_0,Y_1$ be two labeling variables for the two hypotheses respectively that $Y_1=(0_1,\cdots,0_m, 1_1, \cdots, 1_n)$ and $Y_0=(1_1,\cdots,1_{m+n})$. So the CE between $\mathbf{X}$ and $Y_i$ can be calculated as
\begin{equation}
	H_c(\mathbf{X};Y_i) = H_c(\mathbf{X},Y_i)-H_c(\mathbf{X}).
\end{equation}
Then the test statistic for $H_0$ is defined as the difference between the CEs of the two hypotheses, as follows:
\begin{equation}
	T_{ce}=H_c(\mathbf{X},Y_0)-H_c(\mathbf{X},Y_1).
	\label{eq:tce}
\end{equation}
It is easy to know $T_{ce}$ will be a small value if $H_0$ is true and a large value if $H_1$ is true. 

The proposed test statistic can be easily estimated from data by estimating the two terms in \eqref{eq:tce} with the non-parametric estimator of CE. Since the CE estimator is non-parametric, the proposed estimator of the test statistic can be applied to any cases without assumptions. Another merit of the proposed estimator is hyperparameter-free.

\section{Simulation Experiments}
\label{sec:sim}
We evaluate the proposed test statistic with simulated data and compare it with three multivariate non-parametric tests, including the MI-based test \cite{Shekhar2022}, kernel-based test \cite{Gretton2012}, and energy statistics-based test \cite{Szekely2004}. Three simulation experiments are designed to compare them in different situations. The sample size is 500 in all the simulations. The non-parametric CE estimator was also used to estimate MI in the MI-based test. The Gaussian kernel with scale parameter $\delta = 1$ was used for the kernel-based test.

In the first simulation, we generated a sample from bi-variate normal distribution with zero mean $(0,0)$ and variance $\rho = 0.5$. This sample is taken as a reference. We then generated 10 samples with a shifted mean from the zero mean of the reference sample $(i,i),i=0,\cdots,9$, and the same variance of the above normal distribution. The above four two-sample tests were conducted on the reference sample and each newly generated sample with shifted mean. Their test statistics were estimated from the simulated data.

In the second simulation, a reference sample was first generated from a bi-variate normal distribution with zero mean and variance $\rho = 0$. Then 10 samples were generated with the same mean and a variance that increases from $0$ to $0.9$ by step $0.1$, which means an increasing difference with the reference sample. The test statistics of the four two-sample tests were then estimated from the reference sample with each generated sample.

In the third simulation, a reference sample was generated as the same in the second simulation. Then 10 samples were generated with a bi-variate Gaussian copula associated with a standard normal marginal distribution and an exponential marginal distribution with rate as $0.5$. The $\rho$ of the Gaussian copula increases from $0.1$ to $1$ by step $1$, which also means an increasing difference with the reference sample. The test statistics of the four tests were estimated from the reference sample with each sample generated from the Gaussian copula.

\section{Results}
\label{sec:results}

\begin{figure}
	\centering
	\includegraphics[width=\textwidth]{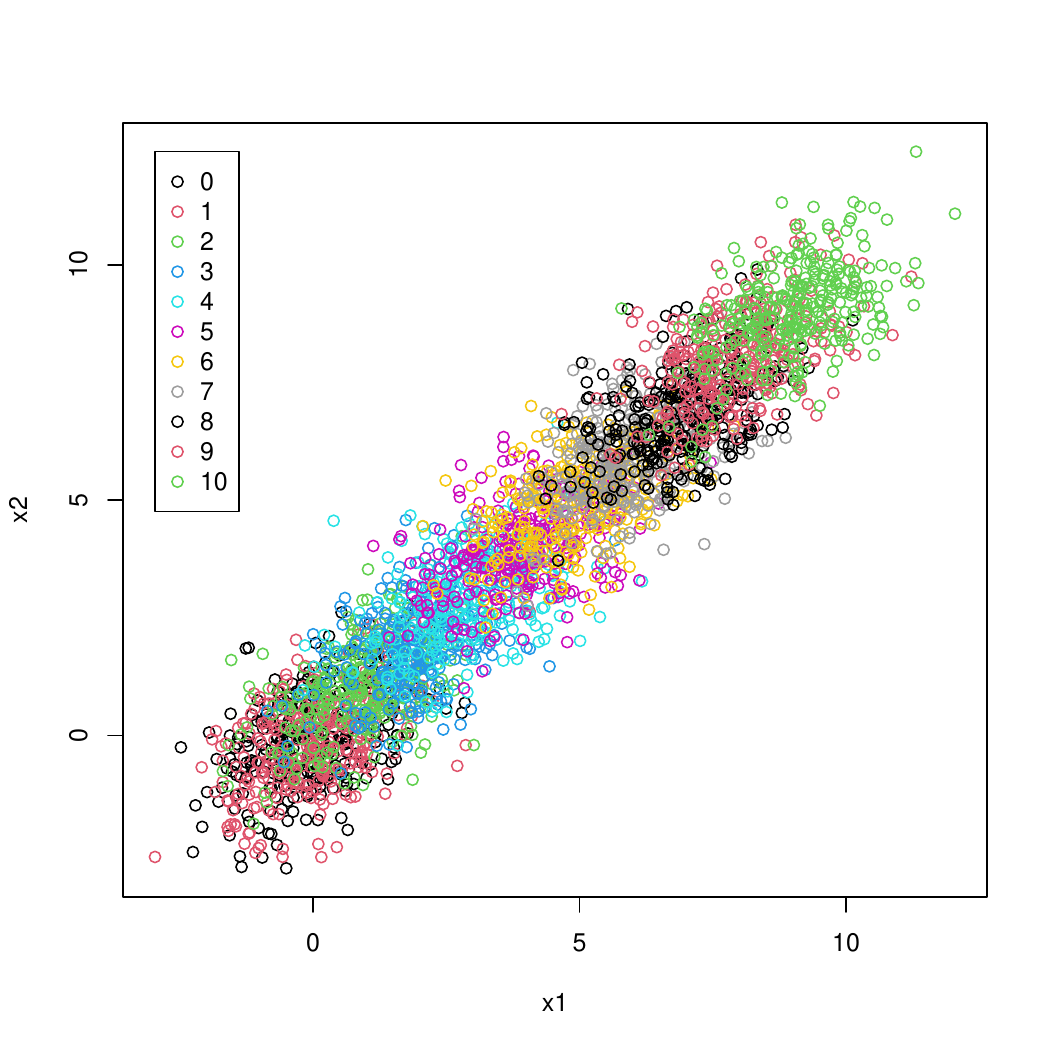}
	\caption{Simulated data. 0 for the reference sample, 1-10 for the newly generated samples.}
	\label{fig:data}
\end{figure}

\begin{figure}
	\centering
	\includegraphics[width=\textwidth]{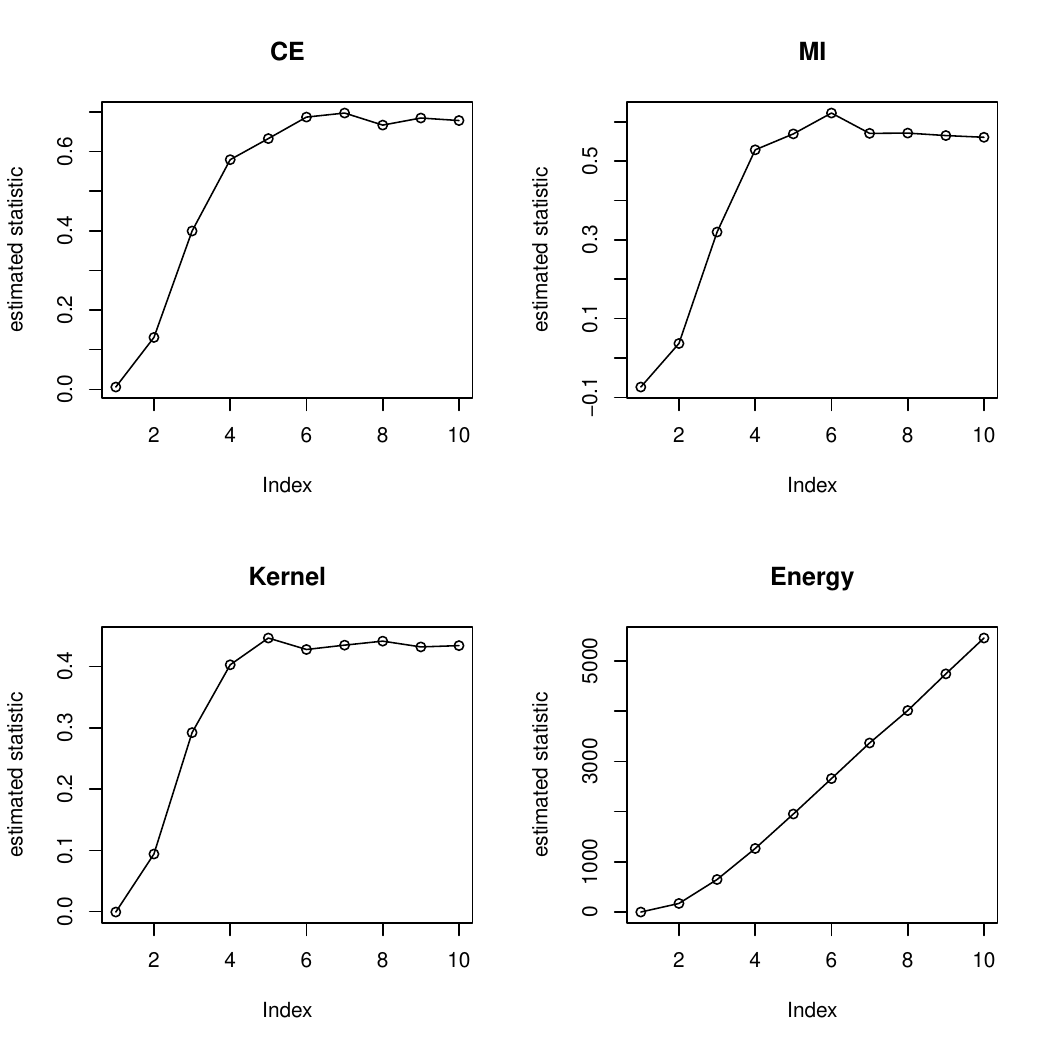}
	\caption{The statistics of the four two-sample tests estimated from the simulated data in the first simulation.}
	\label{fig:sim1}
\end{figure}

The reference sample and the newly generated samples of the first simulation are shown in Figure \ref{fig:data}, from which it can be learned that the generated samples are shifted away from the reference sample gradually. This means that $H_0$ is true at first while is false at last. The estimated test statistics of the four tests are shown in Figure \ref{fig:sim1}. It can be learned from it that the estimated statistic of the proposed test is close to 0 at first and then becomes larger as the mean shifts and remain stable high as the generated sample shifted away from the reference sample. The MI-based test and kernel-based test share similar results while the energy statistics-based test presents a different result.

\begin{figure}
	\centering
	\includegraphics[width=\textwidth]{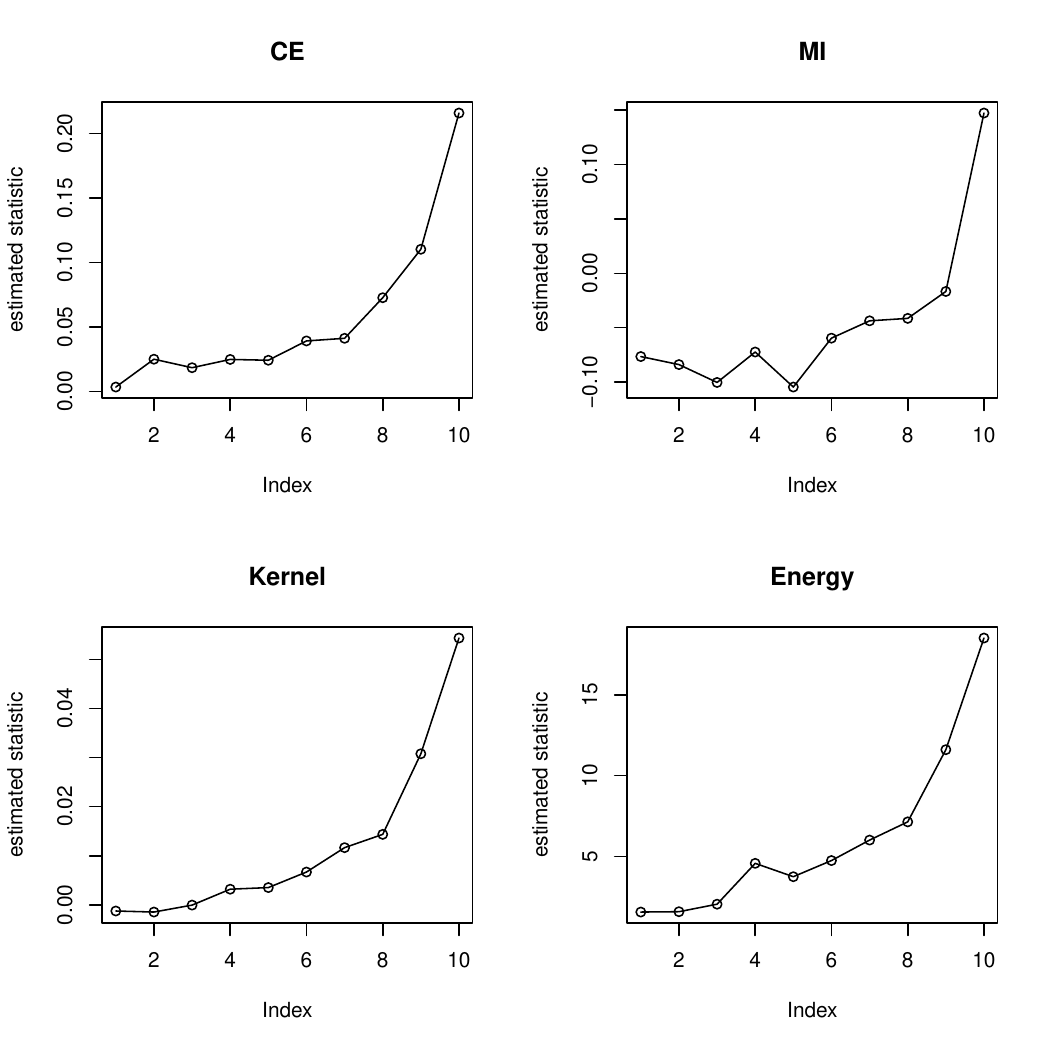}
	\caption{The statistics of the four two-sample tests estimated from the simulated data in the second simulation.}
	\label{fig:sim2}
\end{figure}

The estimated statistics of the four tests in the second simulation are shown in Figure \ref{fig:sim2}. It can be learned from it that the estimated statistics of the four tests are all increased as the $\rho$ of the normal distribution increases. Compared with the proposed test, the MI-based test presents a much unstable result.

\begin{figure}
	\centering
	\includegraphics[width=\textwidth]{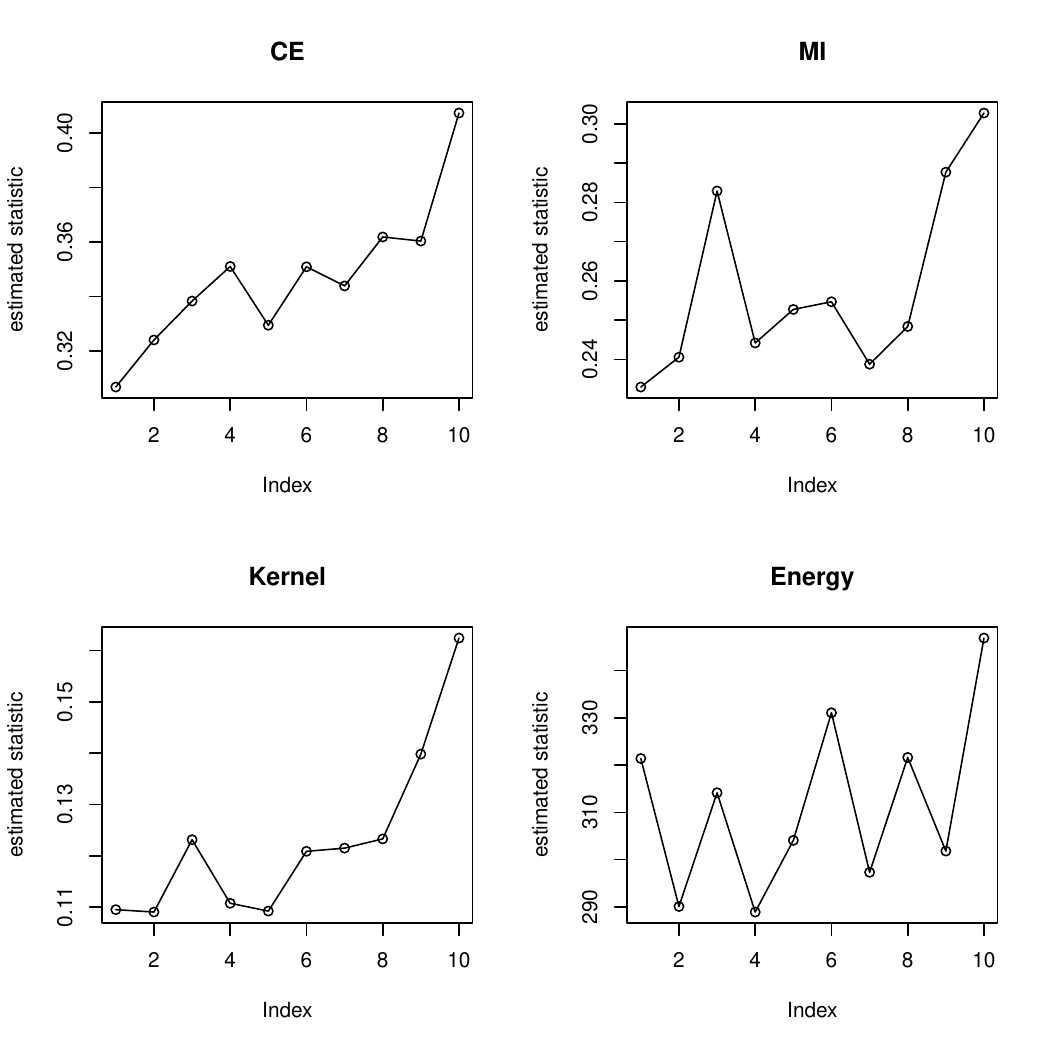}
	\caption{The statistics of the four two-sample tests estimated from the simulated data in the third simulation.}
	\label{fig:sim3}
\end{figure}

The estimated statistics of the four tests in the third simulation are shown in Figure \ref{fig:sim3}. It can be learned from it that the proposed test and the kernel-based test present a result with increasing statistics that can reflect the relationship between the reference sample and the sample generated from the Gaussian copula with increasing covariance parameter while both the MI-based test and the energy statistics-based test fail to present a reasonable result.

\section{Discussion}
\label{sec:discussion}
In this paper we propose a statistic for two-sample test based on CE. This is closely related to the MI-based test since MI is equivalent to negative CE. The difference between ours and the MI-based statistic is that our proposed statistic is defined as the difference between the CEs of the null hypothesis and the alternative while the MI-based statistic is naively defined as only the MI of the null hypothesis. Simulation experiments show that our test can not only present more stable estimation results than the MI-based test but can also present reasonable results in the situations with non-Gaussianity where the MI-based test fails. Simulation results also show that the estimated statistic of our test is close to zero while that of the MI-based test has non-zero bias when $H_0$ is true in the first two simulations.

We compared our test with the kernel-based test and the energy statistics-based test with simulated data. The simulation results show that our test presents similar results as kernel-based test in Gaussian cases while different result with kernel-based test in non-Gaussian cases. As for the energy statistics-based test, it presented a linearly increasing statistic and therefore fails to reflect the stable difference when the newly generated sample has shifted away from the reference sample in the first simulation. It also fails to present a result that reflects the increasing difference between two samples in the third simulation. These mean that our test is competitive with the kernel-based test and better than the energy statistics-based test.

\section{Conclusions}
\label{sec:con}
In this paper we propose a two-sample test based on CE. The proposed test statistic is defined as the difference between the CEs of the null hypothesis and the alternative. The estimator of the test statistic is proposed with the non-parametric estimator of CE. Simulation experiments verify the effectiveness of the proposed test and compare it with three other multivariate non-parametric two-sample tests on three simulated bi-variate normal or bi-variate Gaussian copula data. Experimental results show that the proposed test works well on all three simulations and presents competitive or better performance than the other three tests.

\bibliographystyle{unsrt}
\bibliography{tst}

\end{document}